\documentclass[aps,twocolumn,a4paper,superscriptaddress,nofootinbib]{revtex4-1}
\usepackage{graphicx,subcaption,amsmath,amsfonts,amssymb,multirow,extarrows,bm,acronym,float,caption}
\usepackage[colorlinks,linkcolor=blue,citecolor=blue,urlcolor=blue ]{hyperref}
\floatstyle{plaintop}
\restylefloat{table}
\newcommand{\hwsout}{\textcolor{cyan}{$\mathcal{HT}$: }\bgroup\markoverwith{\textcolor{cyan}{\rule[.5ex]{2pt}{2.5pt}}}\ULon}

\newcommand{\TRC}{TianQin Research Center for Gravitational Physics and School of Physics and Astronomy, Sun Yat-sen University (Zhuhai Campus), Zhuhai 519082, People's Republic of China}

\newcommand{\PMO}{Purple Mountain Observatory, Chinese Academy of Sciences, Nanjing 210023, People's Republic of China}
\newcommand{\USTC}{School of Astronomy and Space Science, University of Science and Technology of China, Hefei, Anhui 230026, People's Republic of China}
\newcommand{\SCUT}{School of Physics and Optoelectronics, South China University of Technology, Guangzhou 510641, People's Republic of China}

\def\apj{Astrophysical Journal}                
\def\apjl{Astrophysical Journal, Letters}             
\def\jcap{Journal of Cosmology and Astropartical Physics}             

\def\mnras{MNRAS}            
\def\prd{Phys.~Rev.~D}       
\def\prl{Phys.~Rev.~Lett}    
\def\pasp{PASP}    

\begin{document}

\title{Quasinormal-modes of the Kerr-Newman black hole: GW150914 and fundamental physics implications}

\author{Hai-Tian Wang}
\affiliation{\PMO}
\affiliation{\USTC}
\affiliation{\TRC}
\author{Shao-Peng Tang}
\affiliation{\PMO}
\affiliation{\USTC}
\author{Peng-Cheng Li}
\email{Corresponding author: lipch2019@pku.edu.cn.}
\affiliation{Center for High Energy Physics, Peking University, No.5 Yiheyuan Rd, Beijing 100871, People's Republic of China}
\affiliation{Department of Physics and State Key Laboratory of Nuclear Physics and Technology, Peking University, No.5 Yiheyuan Rd, Beijing 100871, People's Republic of China}
\affiliation{\SCUT}
\author{Yi-Zhong Fan}
\email{Corresponding author: yzfan@pmo.ac.cn}
\affiliation{\PMO}
\affiliation{\USTC}

\date{\today}

\begin{abstract}
We develop an analytical ringdown waveform model, including both the fundamental and the overtone quasinormal modes, for charged black holes and show that it is precise enough to analyze current gravitational wave data. 
Applying this waveform model to GW150914, the charge to mass ratio of the remnant black hole ($\lambda_{f}$) has been constrained with the sole ringdown gravitational wave data for the first time and the $90\%$ upper limit is $\lambda_{f}\leq 0.38$.
Correspondingly, the deviation parameter of the scalar-tensor vector gravity ($\alpha_{\rm s}$) is limited to be $\alpha_{\rm s}\leq 0.17$. 
Our approach can be directly applied to other binary black hole mergers with loud ringdown radiation. 
\end{abstract}

\maketitle


\acrodef{GW}{gravitational wave}
\acrodef{LIGO}{Laser Interferometer Gravitational-Wave Observatory}
\acrodef{LVC}{LIGO-Virgo collaboration}
\acrodef{BNS}{binary neutron star}
\acrodef{NS}{neutron star}
\acrodef{BH}{black hole}
\acrodef{BBH}{binary black hole}
\acrodef{GR}{general relativity}
\acrodef{PN}{post-Newtonian}
\acrodef{SNR}{signal to noise ratio}
\acrodef{PSD}{power spectral density}
\acrodef{QNMs}{quasinormal modes}
\acrodef{KN}{Kerr-Newman}
\acrodef{PDF}{probability density function}
\acrodef{IMR}{inspiral-merger-ringdown}
\acrodef{GWTC-1}{the first gravitational wave transient catalog}
\acrodef{EM}{electromagnetic}
\acrodef{NR}{numerical relativity}
\acrodef{EHT}{Event Horizon Telescope}
\acrodef{VLBI}{Very Long Baseline Interferometry}

\section{Introduction}\label{sec:intro}
A single distorted black hole is formed after the violent plunge of compact binary black holes. 
The gravitational wave radiation in this stage is called ringdown until it settles to a stationary state. 
The ringdown stage is described by the superposition of quasinormal modes, in the form of damped sinusoids \citep{Schw_PRD_Vishveshwara1970, GW_APJL_Press1971, QNM_APJ_Teukolsky1973}. 
According to the no-hair theorem, the properties of a black hole can be fully described by its mass $M$, specific  spin $\chi$, and charge to mass ratio $\lambda$ \citep{Carter_PRL1971, Israel_PR1967, Israel_CMP1967, Cardoso_IOP2016}, which will be reflected in the frequencies and decay rates of these damped sinusoids. 
Studies of the spectrum of the ringdown signal are also called the black hole spectroscopies and have been developed to extract the properties of the remnant black hole, to test general relativity, the no-hair theorem and the black hole area law \citep{Dreyer:2003bv, Berti:2005ys,Gossan:2011ha,Meidam:2014jpa,Berti:2016lat,PRD_Carullo2019, NoHair_PRL_Isi2019, Overtone_PRX_Giesler2019,2021PhRvL.127a1103I}. 

The astrophysical black holes are widely believed to have negligible charge since the interaction with the surrounding nonelectric neutral environment would quickly neutralize them \citep{Blandford:1977ds}. 
However, up to now unequivocal observational evidence for the neutrality of black hole is still lacking. 
The existing constraints from the electromagnetic observations are model dependent \citep{Zakharov_PRD2014, Zajacek:2018ycb}. 
Additionally, in the minicharged dark matter model, charged black holes could be formed in the early Universe and will not be neutralized by accreting dark matter particles with the opposite charge \citep{Charged_BBH_JCAP_Cardoso2016}. 

In contrast to the electromagnetic observations, gravitational wave observations offer us opportunities to place constraints on the charge of black hole in a more robust way. 
If black holes are indeed charged (electric, magnetic, or dark charge) they can be described by the Einstein-Maxwell theory and will modify the gravitational wave signatures of binary black hole mergers. 
Assuming the charge effect can be introduced as perturbation in the inspiral phase of the gravitational wave waveform,  \citet{2021EPJC...81..769W} have analyzed gravitational wave data, in the first gravitational wave transient catalog, to constrain the charges of the black holes.
Recently, by analyzing gravitational wave data with numerical relativity simulations of the coalescence of the charged binary black holes, \citet{NR_KN_PRL_Bozzola2020} found that GW150914 is compatible with having charge to mass ratio smaller than $0.3$. 
These two works mainly focused on the inspiral phase of gravitational radiation and are limited by the lack of quadrupole-electromagnetic emissions \citep{2021EPJC...81..769W} or the nonspinning assumption of the black holes of GW150914 \citep{NR_KN_PRL_Bozzola2020}. 
In this work we aim to robustly set the first direct bound on the charge of the remnant black hole formed in the mergers with the sole ringdown data. 

For Kerr-Newman black holes, the calculation of the quasinormal modes is rather challenging due to the indissolubility of the coupling between the gravitational perturbations and electromagnetic perturbations \citep{Chandrasekhar:1985kt}. 
So far the exact results are only available in a few specific cases \citep{KNBH_sr_PRL_Pani2013, Pani:2013wsa, QNM_wKN_PRD_Mark2015, QNM_KNBH_PRL_Dias2015,Hod:2014uqa,Hod:2015xlh}. 
The quasinormal modes of slowly-rotating, charged black holes were obtained in \citep{KNBH_sr_PRL_Pani2013, Pani:2013wsa}. 
For weakly charged Kerr-Newman black holes, the fundamental quasinormal modes were studied in \citet{QNM_wKN_PRD_Mark2015}. 
Reference \citep{QNM_KNBH_PRL_Dias2015} successfully obtained the quasinormal modes of Kerr-Newman black holes with numerical calculations, which however were limited to the fundamental modes and  the data are available only for $\chi=\lambda$. 
Alternatively, the quasinormal modes of black holes in general relativity can be approximately calculated via the so-called geodesic correspondence \citep{Ferrari:1984zz,Cardoso:2008bp,Yang:2012he}. 
Though derived only in the eikonal limit, such results can be used for order of magnitude estimates \citep{Charged_BBH_JCAP_Cardoso2016}. 
In this work we propose a modified formula of the geodesic correspondence and calculate both the fundamental and overtone quasinormal modes of the Kerr-Newman black holes. 
The overtones, ignored in the analysis of \citet{Charged_BBH_JCAP_Cardoso2016}, are essential in inferring the final mass and spin magnitude of the remnant with the postinspiral data \citep{Overtone_PRX_Giesler2019}. 
Our results are compared with the dominant fundamental mode of Kerr-Newman given by the numerical relativity simulation \citep{QNM_KNBH_PRL_Dias2015} and the higher modes of the standard Kerr ringdown waveform that is available in the {\sc LALSuite} software library \citep{lalsuite} and the {\sc PyCBC} package \citep{PyCBC_PASP_Biwer2019}. 
The nice consistencies verify the robustness of our Kerr-Newman model. 
Applying our approach to the ringdown data of GW150914 we constrain the charge of the remnant of GW150914 to be $\lambda_f\leq 0.38$ at $90\%$ confidence level. 
Our finding is helpful in testing a specific modified theory of gravity with the ringdown signal. 
In the scalar-tensor-vector gravity theory \citep{Moffat:2005si}, a rotating black hole solution is known to be related to the Kerr-Newman black hole by $\lambda_f=\sqrt{\alpha_s/(1+\alpha_s)}$ \citep{Moffat:2016gkd,Bao:2019kgt}, where $\alpha_s$ can be treated as the parameter characterizing the deviation from general relativity. 
Note that we assume $G = c = 1$ throughout this work unless otherwise specified. 
Hence, the request of $\lambda_f\leq 0.38$ yields a deviation parameter $\alpha_{\rm s}\leq 0.17$.


\section{Method}\label{sec:method}

The quasinormal mode frequency of the Kerr black hole in the eikonal limit ($l\gg1$) reads \citep{Yang:2012he} 
\begin{equation}\label{QNMfre}
\Omega_{lmn}=(l+\frac{1}{2})\left(\Omega_\theta+\frac{m}{l+\frac{1}{2}}\Omega_{prec}\right)-i\beta\gamma_L (n+\frac{1}{2}),
\end{equation}
where $\Omega_\theta$ is the orbital frequency in the polar direction, $\Omega_{prec}$ is the Lense-Thirring precession frequency of the orbit plane, and $\gamma_L$ is the Lyapunov exponent of the orbit, and $n$ is the overtone number. 
Recently, this formula has been shown to be valid for Kerr-Newman black holes \citep{2021PhRvD.104h4044L}. 
The relative error of the result obtained from above formula is larger than $10\%$ \citep{Yang:2012he}, so it cannot be used for quantitative data analysis. 
However, if we replace the prefactor $(l+1/2)$ with $l$ and $\beta=1$ with $(l+l^2+l^3)/(1+l+l^2+l^3)$ in Eq.~(\ref{QNMfre}), the resulting quasinormal mode frequency matches the exact numerical results of both Kerr and Kerr-Newman black holes very well (as shown below). 
Using the formula provided by \citet{Yang:2012he} and \citet{Charged_BBH_JCAP_Cardoso2016}, the explicit expression of the quasinormal mode frequency for $l=m$ can be written as 
\begin{widetext}
\begin{equation}
\begin{aligned}
\Omega_{lln}&=\frac{M \chi  l }{r^2 (2 l +1) \left(r \left(2 \lambda ^2 M^2-3 M r+r^2\right)+M^2 \chi ^2 (M+r)\right)}\Big[2 M r^2 l -2 r^3 l\nonumber\\
&+\sqrt{M^2 \left(\lambda ^2+\chi ^2\right)-2 M r+r^2} \sqrt{-4 \lambda ^2 M^4 \left(\lambda ^2+\chi ^2\right)+8 M^3 r \left(2 \lambda ^2+\chi ^2\right)-\left(4 \lambda ^2+15\right) M^2 r^2+6 M r^3+r^4}\Big]\nonumber\\
&-\frac{i\beta M^{3/2} (2 n+1) r^2 \chi  \sqrt{3 r-4 \lambda ^2 M} \left(M^2 \left(\lambda ^2+\chi ^2\right)-2 M r+r^2\right)}{r^4 \left(r \left(2 \lambda ^2 M^2-3 M r+r^2\right)+M^2 \chi ^2 (M+r)\right)},
\end{aligned}
\label{eq:omega_lln}
\end{equation}
where 
\begin{equation}
\begin{aligned}
r&=2 M (\cos \phi +1)-\frac{\lambda ^2 M \left(\chi ^2+2 \cos (\phi )+2 \cos (2 \phi )\right)}{3 \left(\chi ^2+\cos (\phi )+\cos (2 \phi )\right)}\nonumber\\
&+\frac{\lambda ^4 M}{36 \left(\chi ^2+\cos (\phi )+\cos (2 \phi )\right)^3}\Big[-5 \chi ^4+32 \chi ^2-4 \left(3 \chi ^4-4 \chi ^2-3\right) \cos (\phi )-4\\
& \chi ^2 \cos (4 \phi )+4 \left(\chi ^4+2 \chi ^2+3\right) \cos (2 \phi )+4 \cos (4 \phi )+4 \cos (5 \phi )+4 \cos (6 \phi )\Big]+\mathcal{O}(\lambda^6),\nonumber
\end{aligned}
\label{eq:omega_r}
\end{equation}
with 
\begin{equation}
\phi=\frac{2}{3}\cos^{-1}(-\chi).
\label{eq:omega_phi}
\end{equation}
\end{widetext}
For simplicity, here we only show the result expanded in charge to mass ratio up to quartic order. 

In general relativity, the ringdown waveform of the Kerr-Newman black hole is fully described by two time-dependent polarizations $h(t)=h_{+}(t)-ih_{\times}(t)$, which can be written as 
\begin{equation}
\begin{aligned}
h_{+}(t)-ih_{\times}(t)=&\sum_l^{}\sum_m^{}\sum_n^{N}A_{lmn}\exp\left(i(-\Omega_{lmn}t+\phi_{lmn})\right) \\
&\times{}_{-2}Y_{lm}(\iota),
\end{aligned}
\label{eq:ringdown}
\end{equation}
where $N$ is the total overtone numbers, $A_{lmn}$ ($\phi_{lmn}$) characterize the amplitudes (phases) of each ringdown mode at the peak, $\iota$ is the inclination angle, and ${}_{-2}Y_{m}(\iota)$ are the spin-weighted spherical harmonics \citep{Overtone_PRX_Giesler2019}. 
The response of a single detector $k$ to the gravitational waves is described as 
\begin{equation}
h_{k}(t)=F_{k}^{+}(\alpha, \delta, \psi) h_{+}(t)
+F_{k}^{\times}(\alpha, \delta, \psi) h_{\times}(t),
\end{equation}
where $F_k^{+,\times}(\alpha, \delta, \psi)$ are the antenna beam patterns, $\alpha, \delta$ are the right ascension and declination angles, and $\psi$ is the polarization angle. 

The gravitational wave data stream $d$ is provided by the Gravitational Wave Open Science Center \citep{LIGO_PRX2019}, which contains the signal $h(t)$ and noise $n(t)$. 
In standard gravitational wave data analysis, the detector noise can be assumed to be a Gaussian stochastic process \citep{gw150914_PRL2016}. 
As a Gaussian stochastic process, each set $[n(t_0),n(t_1),,,n(t_{N_s-1})]$ is distributed as multivariate Gaussian probability density function, 
\begin{equation}
\mathbf{n(t)} \sim \mathcal{N}(\boldsymbol{\mu}, \mathbf{\Sigma}),
\end{equation}
where $\boldsymbol{\mu}$ and $\mathbf{\Sigma}_{ij}=\rho(i-j)$ are the mean and the covariance matrix of the noise time series, respectively. 
After applying a high-pass filter with a roll-on frequency of $20$ Hz, the data stream can be treated as zero mean. 
$\rho$ is the autocovariance function 
\begin{equation}
\rho(i-j)=\left\langle n_{i} n_{j}\right\rangle,
\end{equation}
and $\rho(-n)=\rho(N_s-n)$ for $0\leq n< N_s$, where $N_s$ is the total number of samples. 
According to the Wiener-Khinchin theorem, the autocovariance function is the inverse Fourier transform of the \ac{PSD}. 
In our case, the covariance matrix is a circulant Toeplitz matrix and its inverse can be simply solved with {\sc Scipy} package \citep{scipy_Jones2001}. 

For gravitational wave data analysis in time domain, the inner product between two waveforms $h_1(t)$ and $h_2(t)$ can be defined as 
\begin{equation}
(h_1 \mid h_2)=h_1^T\Sigma^{-1}h_2.
\label{eq:inner_p}
\end{equation}
We define the mismatch between two waveforms ($h_1,\,h_2$) as 
\begin{equation}
\begin{aligned}
{\rm Mismatch}&=1-\mathcal{O}(h_1,h_2) \\
&=1-\frac{(h_1 \mid h_2)}{\sqrt{(h_1 \mid h_1)(h_2 \mid h_2)}}.
\end{aligned}
\label{eq:mm}
\end{equation}
Obviously, we have $\rm Mismatch\approx 0$ if these two waveforms match with each other well. 

Given the observed strain series $d_k(t)$ and the gravitational wave signal $h_k(t)$ from waveform model, the log-likelihood function can be written as 
\begin{equation}
\log \mathcal{L}_k=-\frac{1}{2}(d_k-h_k \mid d_k-h_k) +{\rm C},
\label{eq:logl}
\end{equation}
where $\rm C$ is a constant that stands for the normalization term. The log-likelihood function of multiple detectors is the sum of the individual log-likelihoods. 

For GW150914, the ringdown signal in the gravitational wave data is shorter than $0.02$ seconds \citep{gw150914_PRL2016}. 
Our Bayesian analyses are performed with a four seconds data stream around the peak. 
For the ringdown only analyses, there is no waveform data before the peak in the waveform model, thus we add zeros (two seconds) before the peak to match the four seconds of strain data, which also ensures that the data before the peak will not affect the calculation of the likelihood. 
The fundamental mode $(2,2,0)$ and the first overtone mode $(2,2,1)$, which are sufficient for the ringdown analysis \citep{NoHair_PRL_Isi2019, Overtone_PRX_Giesler2019}, have been considered in our analysis. 

To estimate the parameters with the ringdown signal of GW150914, we carry out Bayesian inference with the {\sc Bilby} package \citep{Ashton_APJ2019} and {\sc Dynesty} sampler \citep{Dynesty_MNRAS_Speagle2020}. 
Following \citet{NoHair_PRL_Isi2019}, we fix some extrinsic parameters in our analyses; the geocentric time is set to be $1126259462.408$ GPS, the right ascension is $\alpha=1.95$ rad, the declination is $\delta=-1.27$ rad, the polarization angle is $\psi=0.82$ rad, and the inclination angle is $\iota=\pi$ rad. 
Meanwhile, the parameters above are also fixed when we make a waveform comparison. 
The priors on the final mass $M_f$, the dimensionless spin $\chi_f$, the amplitudes parameter $A_{lmn}$, and the phase parameter $\phi_{lmn}$ are uniformly distributed in the ranges of $[50,\, 100]M_{\odot}$, $[0,\,0.99]$, $[0,\,5\times 10^{-20}]$, and $[0,\,2\pi]$, respectively. 
For the dimensionless charge $\lambda_f$ in the ringdown waveform of the Kerr-Newman black hole, the prior is uniform in the range of $[0,\,1]$. 
Additionally, there is a constraint on $\lambda_f$ and $\chi_f$, i.e., $\lambda_f^2+\chi_f^2\leq 1$, to avoid a naked singularity. 

\section{Results}\label{sec:result}

First and foremost, we compare our Kerr-Newman ringdown waveform with the previous works by considering two detectors of LIGO with the design sensitivity in $O4$ to check whether it can be applied to gravitational wave data analysis. 
We first compare our waveform with the Kerr-Newman ringdown waveform in Ref.~\citep{QNM_KNBH_PRL_Dias2015}, which offers the data of the dominant fundamental mode $(2,2,0)$ in the specific case of $\lambda_f=\chi_f$. 
In this case, the maximum value of the final charge is $\lambda_f=\chi_f\approx 0.7$, which is due to the theoretical constraint mentioned above. 
We use the mismatch defined in Eq.~(\ref{eq:mm}) to investigate whether these two Kerr-Newman ringdown waveform models are highly similar. 
Among the parameters of $(M_f, \lambda_f,\chi_f, A_{220}, \phi_{220})$, $(M_f, \lambda_f,\chi_f)$ are of interest to us. Therefore we assume $A_{220}=0.6,\,\phi_{220}=0,\,\lambda_f=\chi_f$ and calculate the mismatches with $M_f$ lying in the range of $[50,\,200]\,M_{\odot}$ and $\lambda_f=\chi_f$ in the range of $[0,\,0.7]$. 

\begin{figure}
\centering
\includegraphics[width=0.95\linewidth]{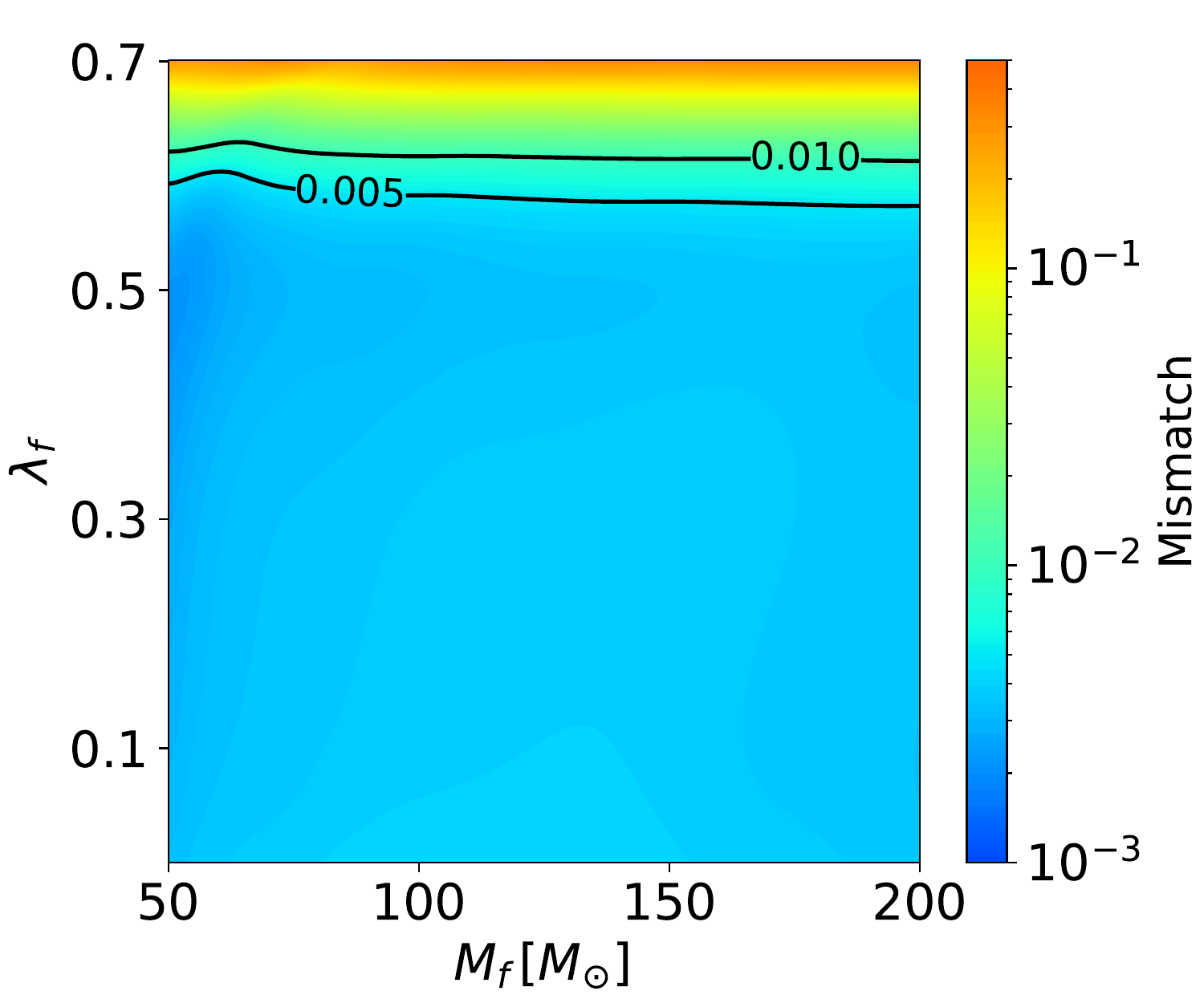}
\caption{
Distribution of mismatch over final charge and final mass, assuming that $\lambda_f=\chi_f$. 
The black lines represent contours on the mismatch between the numerical relativity waveform and our Kerr-Newman waveform. 
}
\label{fig:kn_nr}
\end{figure}

For a signal with \ac{SNR} $\rho$, the mismatch should be smaller than $1/(2\rho^2)$ if two waveform models match well with each other \citep{GW_PRD_Flanagan1998, GW_PRD_Lindblom2008}. 
The network \ac{SNR} for the ringdown stage of GW150914 is about $14$ \citep{NoHair_PRL_Isi2019}, which means that the mismatch threshold is $0.003$. 
Compared with the realistic \ac{PSD}, the \ac{PSD} of design sensitivity for LIGO would overestimate the mismatch between two different waveforms, so $0.005$ is a reasonable choice in our analysis. 
As shown in Fig.~\ref{fig:kn_nr}, the mismatch stays unchanged when $M_f$ ranges from $50\,M_{\odot}$ to $200\,M_{\odot}$. 
When $\lambda_f=\chi_f\leq 0.63$, the mismatch is smaller than $0.01$, and if $\lambda_f=\chi_f\leq 0.60$, it is smaller than the threshold value $0.005$. 
This result indicates that the difference of the dominant fundamental mode between these two waveform models is acceptable for current gravitational wave data analysis (a related analysis was recently given in Ref.~\citep{2021PhRvD.104d4004B}). 

\begin{figure}
\centering
\includegraphics[width=0.95\linewidth]{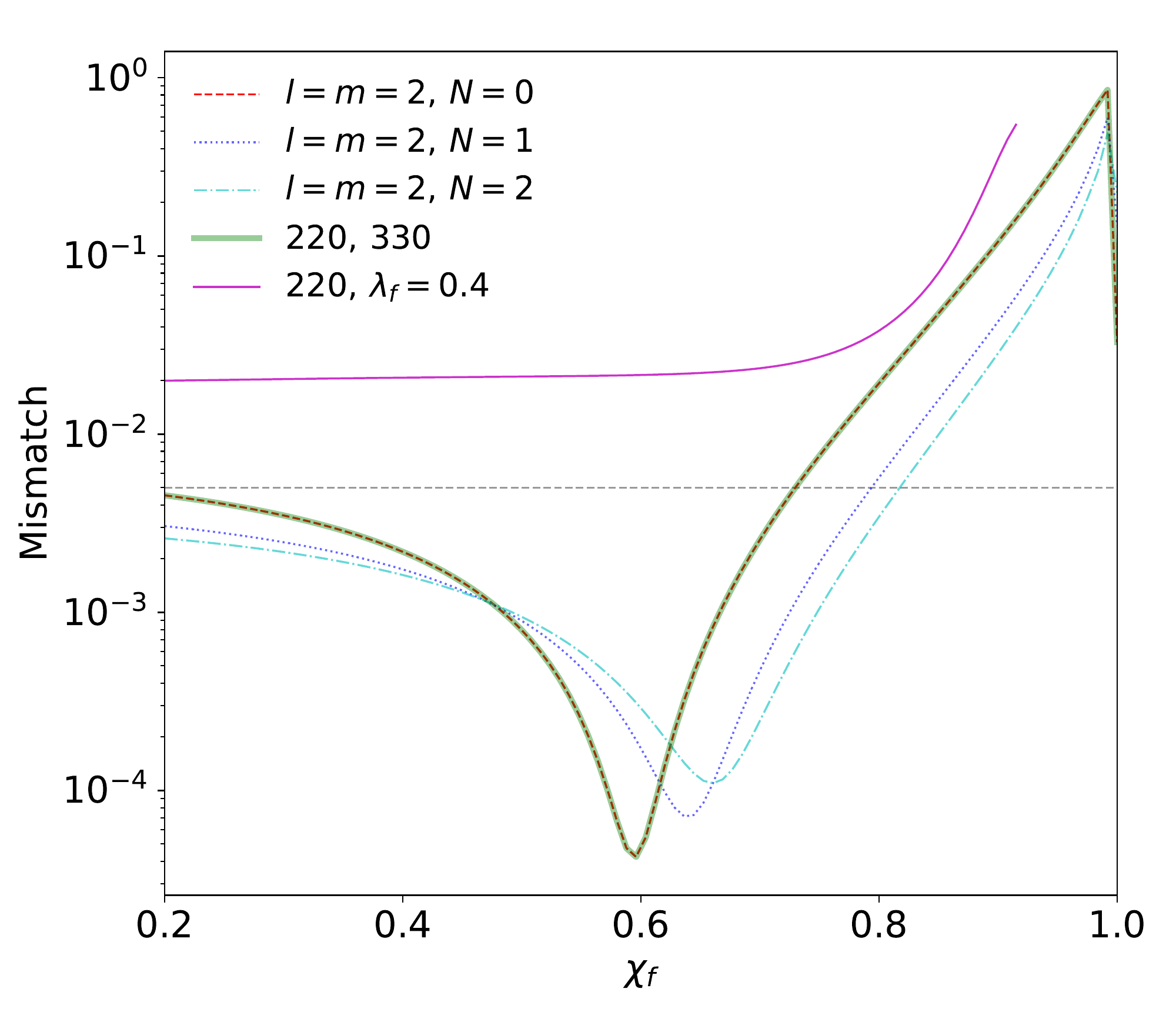}
\caption{
Mismatch ranges over $\chi_f$, assuming $M_f=100 M_{\odot}$. 
The horizontal dashed gray line marks the mismatch at $0.005$. 
We consider the fundamental mode with zero overtone (dashed red), one overtone (dotted blue), two overtones (dash dotted cyan), and higher mode $l=m=3$ (bold green). 
The purple curve represents a special case that $\lambda_f=0.4$ for the fundamental mode in the Kerr-Newman ringdown waveform, other parameters are the same as that in standard Kerr waveform. 
These two ringdown waveform models match better when more overtones are included. 
}
\label{fig:kn_kerr}
\end{figure}

However, when $N>1$ the comparison above can not be made, since there is no available Kerr-Newman ringdown waveform for $N>1$ case up to now. 
Alternatively, by setting $\lambda_f=0$ in our Kerr-Newman ringdown waveform we can make comparison with the standard Kerr ringdown waveform used in Refs.~\citep{NoHair_PRL_Isi2019, Overtone_PRX_Giesler2019}. 
In this case, we assume $M_f=100 M_{\odot}$ since it has little influence on the mismatch. 
The amplitudes $A_{lmn}$ and phases $\phi_{lmn}$ are also fixed to $A_{221}=1.2,A_{222}=0.8,A_{330}=1,\phi_{221}=\phi_{222}=\phi_{330}=0$. 
The final spin ranges from $0$ to $0.998$ and other parameters are the same as in the analysis above. 
It is noticeable that our ringdown waveform behaves better if more overtone modes are included (as shown in Fig.~\ref{fig:kn_kerr}). 
When the final spin is smaller than $0.8$, the mismatch is smaller than $0.005$ if $N\geq 1$. 
Fortunately, the final spins of almost all gravitational wave events are smaller than $0.8$, as shown in Table VIII of \citet{2021PhRvD.103l2002A}. 
We also find that the $(3,3,0)$ mode has little contribution compared with the $(2,2,0)$ mode, which is consistent with the previous study \citep{PRD_Carullo2019}. 
Specially, we compute the mismatch between the Kerr-Newman ringdown waveform ($\lambda_f=0.4$) and the standard Kerr waveform (other parameters are the same for both waveforms). 
Compared with the $\lambda_f=0$ case (dashed red curve in Fig.~\ref{fig:kn_kerr}), the mismatch of $\lambda_f=0.4$ case approaches $2\%$ when $\chi_f\leq 0.8$. 
This means that $\lambda_f=0.4$ leads to a detectable difference for the Kerr-Newman ringdown waveform model. 

In summary, our Kerr-Newman ringdown waveform works very well for the dominant fundamental mode and the cases of $N>0$ have also been checked via the comparison with the standard Kerr ringdown waveform. 
The mismatch can be further reduced by including more overtones in gravitational wave data analysis. 

\begin{figure}
\centering
\includegraphics[width=0.95\linewidth]{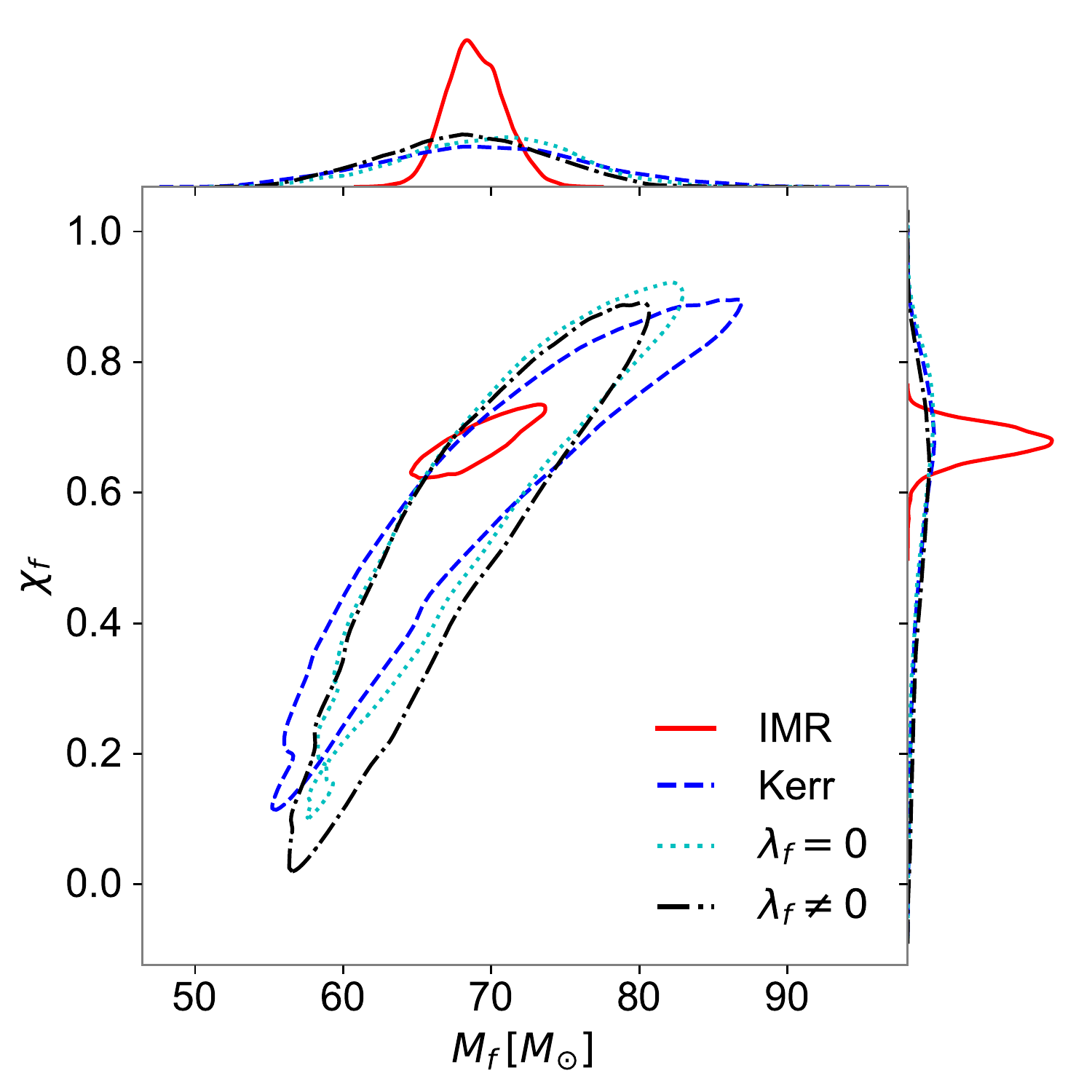}
\caption{
Posterior distributions of final mass and final spin for GW150914. 
All of the contours are $90\%$ credible regions. 
The 1D marginalized posterior distributions for final mass and final spin are shown in the top and right-hand panels respectively. 
The red solid contour represents the result of using IMRPhenomPv2 model \citep{Hannam_PRL2015}, published by \ac{LVC} \citep{LIGO_PRX2019}. 
The analysis using standard Kerr ringdown waveform is represented by dashed blue contour. 
For the Kerr-Newman ringdown waveform model, the distributions are similar for both $\lambda_f=0$ (dotted green) and $\lambda_f\neq 0$ (dash dotted black). 
}
\label{fig:fm_fs_kn}
\end{figure}

\begin{figure}
\centering
\includegraphics[width=0.95\linewidth]{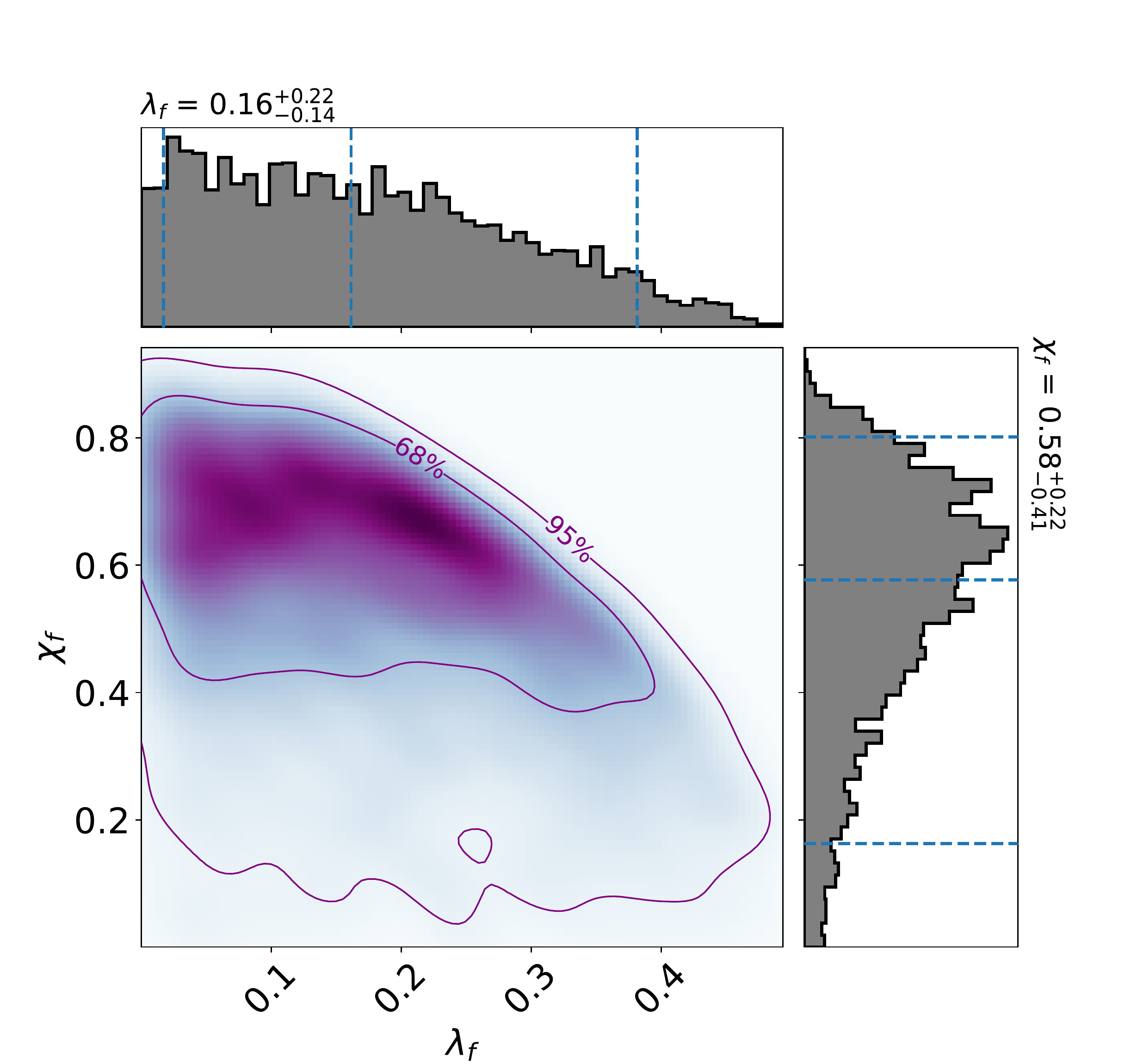}
\caption{
$\lambda_f$ and $\chi_f$ inferred with one overtone, using data starting at peak strain amplitude. 
The purple color map represents the joint posterior distribution. 
The solid curves enclose $68\%$ and $95\%$ of the probability mass. 
The top and right-hand panels show 1D posteriors for $\lambda_f$ and $\chi_f$, respectively. 
The dashed vertical blue lines represent the median value and $90\%$ credible region of $\lambda_f$ and $\chi_f$. 
}
\label{fig:kn_q}
\end{figure}

Using the posteriors of GW150914 (IMRPhenomPv2 case) released by \ac{LVC} \citep{LIGO_PRX2019}, we translate them into the posteriors of the final mass and final spin in the detector frame. 
Then we perform Bayesian analysis with the ringdown signal of GW150914, using standard Kerr ringdown waveform and our Kerr-Newman ringdown waveform. 
For the former we obtain rather similar results as Ref.~\citep{NoHair_PRL_Isi2019}. 
For comparison, we set $\lambda=0$ for the Kerr-Newman ringdown waveform to quantify the difference between it and the standard Kerr ringdown waveform model. 
As shown in Fig.~\ref{fig:fm_fs_kn}, the $90\%$ credible region of IMRPhenomPv2 model is well covered by both models. 
The posterior distributions of $\lambda_f$ and $\chi_f$ obtained by analyzing GW150914 with the Kerr-Newman ringdown waveform are shown in Fig.~\ref{fig:kn_q}. 
Apparently, the distribution of $\lambda_f$ can approach $0$, which is consistent with being uncharged. 
To quantify the difference between results of different waveform models in Fig.~\ref{fig:fm_fs_kn}, we calculate the log-Bayes factors of them, versus the standard Kerr ringdown waveform. 
The log-Bayes factor of $\lambda_f=0$ and $\lambda_f\neq 0$ are $\log_{10}\mathcal{B}^{\lambda_f=0}_{\rm Kerr}=-0.10$, $\log_{10}\mathcal{B}^{\lambda_f\neq 0}_{\rm Kerr}=0.10$, respectively, suggesting no evidence for a strongly charged remnant black hole in GW150914. 

More quantitatively, we have $\lambda_f\leq 0.38$ at $90\%$ confidence level (see Fig.\ref{fig:kn_q}). 
Though a $\lambda \leq 0.3$ was reported in Ref.~\citep{NR_KN_PRL_Bozzola2020}, such a constraint is for the premerger black hole with an additional assumption that only one black hole in GW150914 was charged. 
Therefore we set the first direct bound on the charge of the remnant black hole with the ringdown gravitational wave data. 
Moreover, our analyses were based on the full Bayesian inference, while \citet{NR_KN_PRL_Bozzola2020} fixed all other parameters of the gravitational waveform because of the numerous computations in numerical relativity simulation.

\section{Summary}\label{sec:summary}

In this work, we developed an analytic ringdown waveform model for Kerr-Newman black holes, which is the first one that can be efficiently incorporated in current gravitational wave data analysis. 
The validness of our waveform has been verified through the comparison with other waveforms, including the standard Kerr ringdown waveform 
and the fundamental mode of the Kerr-Newman ringdown waveform with $\chi_{f}=\lambda_{f}$.
We also perform Bayesian analyses with both the Kerr-Newman ringdown waveform and the standard Kerr ringdown waveform. 
The posterior distributions of the final mass/spin of these two waveform models are nicely in agreement with the results from the IMRPhenomPv2 model. 

Applying our approach to GW150914, the final charge of the remnant black hole is constrained to be below $0.38$ (at $90\%$ credibility). 
Correspondingly, the deviation parameter of the scalar-tensor-vector gravity is bounded to be $\alpha_{s}\leq 0.17$. 
It is straightforward to extend our analysis to other events that are suitable for ringdown-only analyses. 
Similar to Ref.~\citep{NR_KN_PRL_Bozzola2020}, since our approach is based on the Einstein-Maxwell theory and the Kerr-Newman metric, it can be directly applied to black holes endowed with hidden or dark charges, and black holes carried with magnetic charge through the electromagnetic duality. 

\begin{acknowledgments}

H.T.W. appreciates Y.M.H., M.Z.H., and J.L.J. for discussions. 
This work has been supported by NSFC under Grants No. 11921003, No. 11847241, No. 11947210, and No. 12047550, and by China Postdoctoral Science Foundation Grant No. 2020M670010. 
The project is supported by Key Laboratory of TianQin Project(Sun Yat-sen University), Ministry of Education. 
We would like to thank https://centra.tecnico.ulisboa.pt/network/grit/files/ringdown/ for the QNM data we used. 
This research has made use of data, software and/or web tools obtained from the Gravitational Wave Open Science Center (https://www.gw-openscience.org), a service of LIGO Laboratory, the LIGO Scientific Collaboration and the Virgo Collaboration. 

\end{acknowledgments}

\bibliographystyle{apsrev4-1}

\bibliography{KNBH_QNM}

\end{document}